\documentclass[12pt]{article}
\usepackage[margin=2cm]{geometry}
\usepackage{comment}
\usepackage{amsmath,amssymb,extarrows,mathtools,graphicx,subfigure,setspace}
\usepackage{cite}
\usepackage{slashed}
\usepackage{color}
\numberwithin{equation}{section}
\usepackage{cite}
\newcommand{\be}{\begin{equation}}
\newcommand{\bea}{\begin{eqnarray}}
\newcommand{\eea}{\end{eqnarray}}
\newcommand{\ba}{\begin{array}}
\newcommand{\ea}{\end{array}}
\newcommand{\ee}{\end{equation}}

\expandafter\ifx\csname mathbbm\endcsname\relax

\else

\fi \textheight 22cm \textwidth 15cm \topmargin 1mm
\begin{document}
\begin{titlepage}
\hfill
\vbox{
    \halign{#\hfil         \cr
           IPM/P-2012/050 \cr
                      } 
      }  
\vspace*{20mm}
\begin{center}
{\Large {\bf  On Holography with Hyperscaling Violation  }\\
}

\vspace*{15mm}
\vspace*{1mm}
{Mohsen Alishahiha$^{a}$, Hossein Yavartanoo$^{b}$ }

 \vspace*{1cm}

{\it ${}^a$ School of physics, Institute for Research in Fundamental Sciences (IPM)\\
P.O. Box 19395-5531, Tehran, Iran \\ }
 \vspace*{1cm}
{\it ${}^b$ Department of Physics, Kyung-Hee University, Seoul 130-701, Korea }
\vspace*{.4cm}

{E-mails: {\tt alishah@ipm.ir},  {\tt yavar@khu.ac.kr}}%

\vspace*{2cm}

\vspace*{2cm}
\end{center}

\begin{abstract}
We study certain features of strongly coupled theories with hyperscaling violation 
by making use of their gravitational duals. We will consider models with  an anisotropic
scaling in time or in one of spatial directions. In particular for the case where 
the anisotropic scaling is along a spatial direction we will compute the holographic
entanglement entropy and show that for specific values of the parameters it exhibits
a logarithmic violation of the area law. We will also probe the backgrounds
 by different
closed and open strings which in turn can be used to read, for example, effective potential of an
external object, drag force and etc.

\end{abstract}

\end{titlepage}

\section{Introduction}

AdS/CFT correspondence \cite{Mal} has thought us to think of a gravitational theory
on an asymptotically locally AdS geometry as a dual description of a strongly coupled field theory 
with a UV fixed point. It is dual in the sense that there is a one to one 
correspondence between objects in the gravitational theory and those in the dual field theory
\cite{{GKP}, {Witten}}. Having an AdS geometry guarantees that the dual field theory has 
conformal symmetry.

It is always challenging to see in what extend the gauge/gravity correspondence can be generalized
to geometries which are not asymptotically AdS. In particular one may consider the case where 
the theory is not conformally invariant, though it may still  be invariant under a certain scaling. Indeed 
such a generalization,  has been made in literature in the context of the application of the  AdS/CFT correspondence in the condensed matter physics (for a review see for example\cite{Hartnoll:2009sz}). 
A prototype example  of such a model is a theory with the Lifshitz fixed point where the theory
has a spatial isotropic scale invariance  which is characterized by a dynamical exponent, $ z$, as follows 
\be
t\rightarrow \lambda^z t,\;\;\;\;\;\;\;\;\;\;x_i\rightarrow \lambda x_i,
\ee
where $t$ is time and $x_i$’s are spatial directions of the space-time.

The holographic description of a $(d+1)$-dimensional theory with the Lifshitz fixed point has been considered 
in\cite{KLM}\footnote{See also \cite{Koroteev:2007yp} for an earlier work on a geometry with the Lifshitz scaling.} where it was proposed  that the corresponding gravitational dual can be defined by a gravity 
on the following metric (see also \cite{Taylor})\footnote{Geometries with non-relativistic conformal
symmetry have also been considered in \cite{{Son},{BMc}}. For earlier work on geometry 
with Schr\"odinger symmetry see \cite{Alishahiha:2003ru}.}
\be\label{metric0}
ds^2_{d+2}=R^2\left(-\frac{dt^2}{r^{2z}}+\frac{dr^2}{r^2}+\frac{\sum_{i=1}^ddx_i^2}{r^2} \right),
\ee
where $R$ is the radius of curvature\footnote{Through out this paper we set $R=1$.}.

It is worth to mention that a metric with Lifshitz isometry is not a solution of a pure cosmological Einstein gravity. This is simply because in the pure Einstein gravity there is nothing to produce an anisotropic in 
the space-time. In fact to 
obtain  such a solution one needs to couple the Einstein gravity  to other fields. In the minimal case 
the extra field could be a massive gauge field \cite{Taylor} which can, indeed, produce an anisotropic in 
the space-time leading to Lifshitz geometry. 

More naturally the Lifshitz metric may be found in an Einstein-Maxwell-Dilaton theory\cite{{Eric}, {Peet}, {BBPZ}, {GKPTIW}, {Sandip}, {Berglund}}. Actually by including both dilaton (in general a scalar with non-trivial
potential) and an abelian gauge field, one can find even more interesting metrics. In particular one may
obtain a metric with the following form \cite{Kiritsis} 
\be
\label{metric1}
ds_{d+2}^2=r^{\frac{2\theta}{d}} \left(-\frac{dt^2}{r^{2z}}+\frac{dr^2}{r^2}+\frac{\sum_{i=1}^ddx_i^2}{r^2} \right),
\ee
where the constants $z$ and $\theta$ are  dynamical and  hyperscaling violation exponents, respectively. This is the most general geometry which is spatially homogeneous and covariant under the following scale transformations
\be
t\rightarrow \lambda^z t, \quad r\rightarrow \lambda r, \quad x_i\rightarrow \lambda x_i, \quad  
ds_{d+2} \rightarrow \lambda^{\frac{\theta}{d}} ds_{d+2}.
\ee
Note that the metric \eqref{metric1} is conformally  a Lifshitz geometry which is the
scale invariant limit $\theta=0$ of \eqref{metric1}.  Indeed with a non-zero $\theta$, the distance is not
invariant under the scaling which in the context of AdS/CFT indicates  violations of 
hyperscaling in the dual field theory. More precisely, while in $(d+1)$-dimensional theories 
without  hyperscaling (dual to background\eqref{metric0})  the entropy  
scales as $T^{d/z}$ with temperature, in the present case (dual to background \eqref{metric1}) it scales
as $T^{(d-\theta)/z}$ \cite{{Gouteraux:2011ce},{Huijse:2011ef}}. 

Holographic aspects of theories with  hyperscaling violation have been studied in\cite{Dong:2012se}
where the authors have shown that  in order to have a physically sensible dual field theory one should 
assume
\be
 (d-\theta)(dz-\theta-d) \geq 0,\quad      (z-1)(d+z-\theta) \geq 0.
\ee 
which is a necessary condition  for the null energy condition to be satisfied, {\it i.e.}  $T_{\mu\nu} N^{\mu}N^{\nu}\geq 0$. In particular it was shown that for $\theta=d-1$ the holographic entanglement 
entropy shows a logarithmic violation of the area law indicating that the dual theory exhibits an 
${\cal O}(N^2)$ 
Fermi surface \cite{{Ogawa:2011bz},{Huijse:2011ef}}. Therefore the geometry \eqref{metric1}
with specific values of its parameters could provide a gravitational dual for a system with 
Fermi surface in any dimensions.  

However, we note that due to hyperscaling violation as well as the behavior of dilaton field 
at large distances, the background \eqref{metric1} cannot provide a dual description
of a theory in all range of energies from UV to IR \cite{{Bhattacharya:2012zu},{Kundu:2012jn}}
(see also\cite{Dong:2012se}) .  In fact gravity on the background \eqref{metric1} may be considered 
as an effective theory valid over an intermediate energy scale.

One of the aim of the  present paper is to further  explore certain features of hyperscaling 
violation theories by 
making use of the gravity dual given by \eqref{metric1}.  More precisely we will probe the background
by an open string which in turns can be used to study the response of the theory to an 
external object. In particular we shall study potential between external objects, drag force as well
as a rotating object. We will also make a comment on a closed string probing the background.

Following \cite{deBoer:2011wk} we will also consider a geometry with an  anisotropic scaling 
in a spatial direction. The corresponding geometry can be obtained  by  
a double Wick rotation from  the metric \eqref{metric1}.
To be more concrete,  consider a double Wick rotation as follows
\be
t\rightarrow i y,\;\;\;\;\;\;\;\;\;\;\;x_{d}\rightarrow i t,
\ee
then the metric \eqref{metric1} becomes
\be
\label{metric2}
ds_{d+2}^2=r^{\frac{2\theta}{d}} \left(\frac{dy^2}{r^{2z}}+\frac{dr^2}{r^2}+\frac{\sum_{i=1}^{d-1}dx_i^2
}{r^2}-\frac{dt^2}{r^2} \right).
\ee
The above metric is covariant under the following scale transformations
\be
y\rightarrow \lambda^z y,\;\;\;\;\;\;r\rightarrow \lambda r,\;\;\;\;\;\;ds_{d+2}\rightarrow 
\lambda^{\frac{\theta}{d}} ds_{d+2},\;\;\;\;\;\;(t,x_i)\rightarrow \lambda (t,x_i),
\ee
for $i=1,\cdots,d-1$.

In the present paper we will also intend to study holographic aspects a theory whose dual
gravitational theory is given by  a
gravity on the background \eqref{metric2}. In particular, we study holographic entanglement
entropy in this background. We show that for $\theta=d-1$ for any $z$, the  entanglement 
entropy exhibits a  logarithmic violation of the area law showing that the geometry may also 
provide a suitable gravitational description for theories with ${\cal O}(N^2)$ Fermi surface.  
We will also probe this geometry by an open string.

The paper is organized as follows. In the next section we study holographic entanglement 
entropy using the metric \eqref{metric2}. In section three we will consider an open string probing 
the metrics \eqref{metric1} and \eqref{metric2}. For completeness of our study we will consider 
a closed string probe in section four. The last section is devoted to discussions.
In what follows we only consider the case with $\theta\leq d$.


\section{Holographic entanglement entropy}

In this section we will study the entanglement entropy by making use of the AdS/CFT correspondence. To compute the entanglement entropy via AdS/CFT correspondence
one needs to minimize a surface in the bulk gravity.  More precisely, given a gravitational  theory with
 the  bulk Newton's constant $G_N$, the holographic entanglement entropy 
is given by \cite{RT:2006PRL,RT:2006}
\be\label{EE}
S_A=\frac{\mathrm{Area}(\gamma_A)}{4G_N},
\ee
where $\gamma_A$ is the  minimal surface in the bulk whose boundary coincides with the boundary of the entangling region.

 It is important to note that in order to compute the holographic entropy, since we are dealing 
with a spatial surface,  one needs to work  at  constant time. In other words, the $g_{tt}$ component
of the metric does not, directly,  contribute to the holographic entanglement entropy. On the other hand, 
for the hyperscaling violating geometry \eqref{metric1}, there are non-trivial effects due to 
non-relativistic nature of the metric which in turn comes from the $g_{tt}$ component. 

Holographic entanglement entropy of the background \eqref{metric1} has been
studied in \cite{Dong:2012se} where it was shown that the entanglement entropy exhibits interesting behaviors
for $d-1\leq \theta\leq d$. Indeed while the theory has extension entropy for $\theta=d$ it shows
logarithmic violation of area law for $\theta=d-1$. 

In this section following \cite{deBoer:2011wk}  we would like to study holographic entanglement 
entropy for the background \eqref{metric2}. For this metric even at constant time slices the metric 
has an anisotropic scaling.  

To proceed,  we will calculate entanglement entropy for a strip subsystem in the dual theory.
From gravity point of view  one needs to minimize a surface in the geometry \eqref{metric2}  whose intersection with the boundary 
coincides to the strip.  In course to do so, we recognize two possible ways to 
embed the strip in the theory depending on the directions we choose for the strip.
In the first  case the width of the strip is along $y$ direction while in the second case one may pick up one 
of the isotropic directions  $x_i$'s for the width of the strip. In what follows we consider both of them. 

\subsection*{Case A}

For the first case, where the width of strip is along the anisotropic scaling direction, consider the following strip 
\be\label{stripA}
\frac{\ell}{2}\leq y\leq \frac{\ell}{2}, \;\;\;\;\;\;\;\;\;\;\;\;\;0\leq x_i\leq L,\;\;{\rm for}\;i=1,\cdots,d-1.
\ee 
Now the aim is to  minimize a surface on the bulk whose boundary is the above strip and its profile 
in the bulk is given by $y=y(r)$. From the geometry \eqref{metric2} the induced metric on this surface reads
\be
ds_{ind}^2=r^{2\frac{\theta}{d}}\left[ \left(\frac{y'^2}{r^{2z}}+\frac{1}{r^2}\right)dr^2+\frac{\sum_{i=1}^{d-1}dx_i^2
}{r^2} \right].
\ee
Therefore the area of the surface is
\be
A=L^{d-1}\int_\epsilon dr\; \frac{\sqrt{r^{2(z-1)}+y'^2}}{r^{d+z-\theta-1}}.
\ee
where prime represents derivative with respect to $r$.
Note that since we are considering $\theta\leq d$, one  has $d+z-\theta-1\geq 0$.
Therefore the area always diverges at UV where $r\rightarrow 0$, and thus we had to introduce
a UV cut off $\epsilon$ in the above expression.

To minimize the area one may consider the above expression as an action of  a one dimensional mechanical
 system whose momentum is a constant of motion
\be
\frac{1}{r^{d+z-\theta-1}}
\frac{y'}{\sqrt{r^{2(z-1)}+y'^2}}=\frac{1}{r_0^{d+z-\theta-1}}={\rm constant}.
\ee
Here $r_0$ is the turning point where $y'|_{r_0}\rightarrow\infty$. The above constraint can be solved
to find the  width of the strip  as a function of turning point $r_0$,
\be
l=2r_0^z\int_0^1d\xi\;\frac{\xi^{d+2z-\theta-2}}{\sqrt{1-\xi^{2(d+z-\theta-1)}}}.
\ee
Finally for the area we arrive at 
\be
A=\frac{L^{d-1}}{r_0^{d-\theta-1}}\int_{\frac{\epsilon}{r_0}}^1d\xi \;
\frac{\xi^{\theta-d}}{\sqrt{1-\xi^{2(d+z-\theta-1)}}},
\ee
It is now possible to preform the above integrals and eliminate
$r_0$ between them to find the  entanglement entropy as a function of width $\ell$. 
For $\theta\neq d-1$,  one finds
\be
\ell=c_0 r_0^z,\;\;\;\;\;\;\;\;\;A=\frac{L^{d-1}}{d-\theta-1}\left(\frac{1}{\epsilon^{d-\theta-1}}-
\frac{b_0}{r_0^{d-\theta-1}}\right),
\ee
with 
\be
c_0=\frac{2\sqrt{\pi}}{z}\;\frac{\Gamma\left(\frac{d+2z-\theta-1}{2(d+z-\theta-1)}\right)}
{\Gamma\left(\frac{z}{2(d+z-\theta-1)}\right)},\;\;\;\;\;\;\;\;\;\;\;
b_0=\sqrt{\pi}\;\frac{\Gamma\left(\frac{-d+\theta+1}{2(d+z-\theta-1)}\right)}
{\Gamma\left(\frac{z}{2(d+z-\theta-1)}\right)}.
\ee
Plugging the obtained minimized area in the equation \eqref{EE}, one can find  the entanglement 
entropy as follows
\be 
S=\frac{L^{d-1}}{4(d-\theta-1)G_N}\left(\frac{1}{\epsilon^{d-\theta-1}}-b_0
\frac{c_0^{(d-\theta-1)/z}}{\ell^{(d-\theta-1)/z}}\right).
\ee 
As we already mentioned,  one  would not expect that the metric provides the gravitational dual
for whole range of the energies. In fact above a given dynamical scale, $r_F$, one would expect that 
the UV theory will be described by a UV completion geometry\cite{Dong:2012se}\footnote{
With the dynamical scale $r_F$ the metric \eqref{metric1} should be read$$
ds_{d+2}^2=\left(\frac{r}{r_F}\right)^{\frac{2\theta}{d}} \left(-\frac{dt^2}{r^{2z}}+\frac{dr^2}{r^2}+\frac{\sum_{i=1}^ddx_i^2}{r^2} \right)$$.}.
 Taking into account 
the dynamical scale $r_F$, the entanglement entropy reads
\be 
S=\frac{L^{d-1}}{4(d-\theta-1)G_N}\left[\left(\frac{\epsilon}{r_F}\right)^\theta\;\frac{1}{\epsilon^{d-1}}-
\left(\frac{\ell^{1/z}}{r_F}\right)^\theta\;
\frac{b_0c_0^{(d-\theta-1)/z}}{\ell^{(d-1)/z}}\right].
\ee 

On the other hand for  $\theta=d-1$ and any $z$, one gets $z\ell =2r_0^z$ while for the area one finds
$A=-L^{d-1}\ln\frac{\epsilon}{2^{1/z}r_0}$. Therefore in this case, considering the 
scaling $r_F$,   one arrives at
\be
S=\frac{1}{4z\pi G_N}\;\frac{L^{d-1}}{r_F^{d-1}}\;\ln\frac{z\ell}{\epsilon^z},
\ee
showing that the dual theory exhibits logarithmic violation for the area law which could be an 
indicator of having a Fermi surface in the model. Note also that  for $d=\theta$, the expression for $A$ is finite and indeed for the entropy we find  
\be 
S\sim L^{d-1}\ell^{1/z}.
\ee 

It is interesting to note that, even though in the present case where the anisotropic scaling 
was along a spatial direction, the system exhibits similar behaviors as that in \cite{Dong:2012se}
where the anisotropic scaling was along the time direction. Actually the results are almost
identical up to a $1/z$ factor in the power which can be understood from the fact that 
the scaling of the width of the strip  in these two cases are related by factor of $z$. Therefore 
from the results of  \cite{Dong:2012se} one can find our results by just replacing 
$r_0\rightarrow r_0^{z}$, or $\ell\rightarrow \ell^{1/z}$.

\subsection*{Case B}

Let us consider  a case where the width of strip is along one of $x_i$'s directions. The corresponding
strip subsystem may be given by 
\be\label{stripB}
\frac{\ell}{2}\leq x_{d-1}\leq \frac{\ell}{2},\;\;\;\;\;\;0\leq y\leq L,\;\;\;\;\;\;0\leq x_i\leq L,\;\;\;\;
{\rm for}\;i=1,\cdots, d-2.
\ee 
The holographic entanglement entropy is given by a minimal surface whose shape at the boundary coincides
to the above stripe and  has a profile in the bulk given by $x_1=x(r)$.  The induced 
metric on the surface is
\be
ds^2_{ind}=r^{2\frac{\theta}{d}}\left[\frac{dy^2}{r^{z}}+(1+x'^2)\frac{dr^2}{r^2}+
\frac{\sum_{i=1}^{d-2}dx_i^2}{r^2}\right].
\ee
Thus the area reads
\be
A=L^{d-1}\int dr\;\frac{\sqrt{1+x'^2}}{r^{d+z-\theta-1}}.
\ee
One can  go through the well known procedure to minimize the surface, as we did in the previous
case. Doing so, one arrives at the following expressions for the width $\ell$ and area $A$
\be
\ell=2r_0\int_0^1d\xi\;\frac{\xi^{d+z-\theta-1}}{\sqrt{1-\xi^{2(d+z-\theta-1)}}},\;\;\;
A=\frac{L^{d-1}}{r^{d+z-\theta-2}}\int_{\frac{\epsilon}{r_0}}^1d\xi\;
\frac{\xi^{-(d+z-\theta-1)}}{\sqrt{1-\xi^{2(d+z-\theta-1)}}}
\ee
Then it is easy to calculate  the above integrals to find the width,$\ell$,  and area $A$ as functions of turning 
point $r_0$. By eliminating $r_0$ between these two functions, we will find the holographic
 entanglement entropy as a function of  width $\ell$. In fact for $\theta\neq d+z-2$ one finds
\be
\ell=c_0 r_0,\;\;\;\;\;\;\;\;\;\;\;A=\frac{L^{d-1}}{d+z-\theta-2}\;\left(\frac{1}{\epsilon^{d+z-\theta-2}}-
\frac{b_0}{r^{d+z-\theta-2}}\right),
\ee
where
\be
c_0=\sqrt{\pi }\;\frac{{\Gamma}\left(\frac{d+z-\theta }{2 (d+z-\theta-1 )}\right)}{{\Gamma}\left(\frac{1}{2 (d+z-\theta-1 )}\right)} ,\;\;\;\;\;\;b_0=\sqrt{\pi }\;\frac{ {\Gamma}\left(\frac{d+z-\theta }{2 (d+z-\theta-1 )}\right)}{ {\Gamma}\left(\frac{1}{2 (d+z-\theta-1 )}\right)}.
\ee
Therefore, taking into account the dynamical scale $r_F$, the entanglement entropy is
\be
S=\frac{L^{d-1}}{4(d+z-\theta-2)G_N}\;\left[\left(\frac{\epsilon}{r_F}\right)^\theta
\frac{1}{\epsilon^{d+z-2}}-\left(\frac{l}{r_F}\right)^\theta
\frac{b_0c_0^{d+z-\theta-2}}{\ell^{d+z-2}}\right].
\ee

On the other hand for $\theta= d+z-2$ one gets $\ell=2r_0$, while the area exhibits a logarithmic 
behavior, $A=L^{d-1}\ln\frac{2r_0}{\epsilon}$. Thus the entanglement entropy is found to be
\be
S=\frac{1}{4G_N}\;\frac{L^{d-1}}{r_F^{d+z-2}}\;\ln\frac{\ell}{\epsilon},
\ee
that  violates the area law. Here we have also restored the dynamical scale $r_F$. If we restrict 
ourselves to the case of $z\geq 2$ we get the above logarithmic violation of area law just 
for $d=\theta$. Thus, taking into account that we are  interested in $\theta\leq d$, the subsystem defined by \eqref{stripB},  unlike the case of \eqref{stripA}, can never have an extensive entropy.  

However we note that, if we let the dynamical exponent to be one, $z=1$, then the 
entanglement entropy of the subsystem \eqref{stripB} exhibits  logarithmic violation of area
for $\theta=d-1$. Moreover for $z=1$ one may have  $\theta=d+z-1$  which means
$\theta=d$. In this case the area turns out to  be a constant leading to
an extensive  entanglement entropy as follows
\be
 S\sim L^{d-1}\ell.
\ee
Of course it is worth mentioning that in this case the metric \eqref{metric1} is, actually, flat space
(more precisely, it is a metric of upper half space) and it is not clear what the 
holographic entanglement entropy means!


\section{Open string probe}

In this section we would like to probe the background \eqref{metric1} and \eqref{metric2} by an
open string. As we already mentioned these geometries are expected to provide gravitational
dual for theories with hyperscaling violation in a specific range of energies. Indeed these 
geometries may be though of as an effective holographic description of the dual theories with 
a UV cut off set by the dynamical scale $r_F$ over which the geometries may not provide 
a good description.

Therefore by probing these geometries with an open string one can explore the interaction of an 
external object in an effective theory exhibiting hyperscaling violation. In particular
we will consider a static or moving open string which in turn can be used to find the effective
potential between two external point like objects  or the drag force the external object may experience.

\subsection{Wilson loop}

The effective potential between external point like objects may be obtained by computing 
the Wilson loop in the model. Actually AdS/CFT correspondence has provided a simple 
prescription to compute the Wilson loop in a strongly coupled field theory by making use
of its gravitational dual\cite{{Rey:1998ik},{Maldacena:1998im}}. In this  
context the Wilson loop can be computed by minimizing a world sheet of an open string whose ends  
on the boundary are, indeed, representing the external objects.

To proceed consider the following ansatz for the open string
\be
t=\tau,\;\;\;\;\;\;\;r=\sigma,\;\;\;\;\;\; x_1=x(r),
\ee
one the background \eqref{metric1}.  For this ansatz the Nambu-Goto  action reads
\be
I=-\frac{1}{2\pi \alpha'}\int dt dr\;{r^{-\frac{d(z+1)-2\theta}{d}}} {\sqrt{1+{x'}^2}}.
\ee 
This action can be thought of as a one dimensional mechanical system for $x$ whose momentum is a 
constant of motion. With this interpretation, denoting the turning  point by $r_0$  where $x'\rightarrow \infty$, one finds
\be
\ell=2\int_0^{r_0} dr \;\frac{\left(\frac{r}{r_0}\right)^{\frac{d(z+1)-2\theta}{d}}}{\sqrt{1- \left(\frac{r^2}{r_0^2}\right)^{\frac{d(z+1)-2\theta}{d}}}},
\ee   
where $\ell$ is the distance between two external objects.  Note that in the cases we are 
interested in one has $d(z+1)-2\theta\geq 0$. It is, then, easy to perform the integral yielding to
 \be
\ell=c_0 r_0, \;\;\;\;\;\;\;{\rm  with}\; c_0=\sqrt{\pi}\frac{ {\Gamma}\left(\frac{d (2+z)-2 \theta }{2 (d(z+1) -2 \theta )}\right)}{{\Gamma}\left(\frac{d}{2( d( z+1)-2 \theta) }\right)}
\ee
On the other hand  the on shell action of the string reads 
\be
I=-\frac{T}{2\pi\alpha'}\int_\epsilon^{r_0}dr \; \frac{ r^{-\frac{d(z+1)-2\theta}{d}}}{\sqrt{1- \left(\frac{r^2}{r_0^2}\right)^{\frac{d(z+1)-2\theta}{d}}}}=-\frac{Td}{2\pi(zd-2{\theta})\alpha'}
\left(\frac{1}{\epsilon^{z-\frac{2 \theta }{d}}}
+\frac{I_0}{ r_0^{z-\frac{2 \theta }{d}}}\right),
\ee
where $T$ is the time interval over which the Wilson loop is calculated, and 
\be
I_0=\frac{\sqrt{\pi }}{1+2  z-4\frac{ \theta}{d} }\;
\frac{ {\Gamma}\left(\frac{-d z+2 \theta }{2 (d( z+1)-2 \theta )}\right)}
{ {\Gamma}\left(\frac{-d(1+2z)+4\theta}{2( d( z+1)-2 \theta) }\right)}
\ee
Therefore for $zd\neq 2\theta$, restoring the dynamical scale $r_F$, the effective potential reads
\be
V_{\rm eff}= -\frac{d}{2\pi(zd-2{\theta})\alpha'}
\left[\left(\frac{\epsilon}{r_F}\right)^{\frac{2 \theta }{d}}\frac{1}{\epsilon^{z}}
+\left(\frac{\ell}{r_F}\right)^{\frac{2 \theta }{d}}
\frac{I_0c_0^{z-\frac{2 \theta }{d}}}{ \ell^{z}}\right].
\ee
While for $zd=2\theta$ one has $\ell=2r_0$ and  for the action one gets
\be
I=-\frac{T}{2\pi\alpha'}\int_{\frac{\epsilon}{r_0}}^1\frac{d\xi}{\xi\sqrt{1-\xi^2}}=-\frac{T}{2\pi\alpha'}
\ln\frac{2r_0}{\epsilon}.
\ee
Thus in this case the effective potential has a logarithmic behavior  as follows
\be
V_{\rm eff}=-\frac{1}{2\pi r_F^z\alpha'}\ln\frac{\ell}{\epsilon}.
\ee

So far we have considered an open string probing the metric \eqref{metric1}. We would like to
extend our study for the metric \eqref{metric2} too. In this case we recognize two distinctive cases
depending on whether the distance between the external objects is along the
anisotropic scaling direction, $y$, or along one of  $x_i$'s directions. Actually for both of cases
one may go through the similar computations as presented above to calculate the 
effective potential between two point like external objects. Doing so one arrives at
\bea
{\rm Along}\;y \;\;{\rm direction}\;&& V_{\rm eff}\sim \left\{\ba{ll} -\ell^{\frac{2\theta-d}{zd}}& \theta\neq \frac{d}{2},\cr &\cr -\frac{1}{z}\ln\frac{\ell}{\epsilon} &\theta=\frac{d}{2},\ea\right.\cr &&\cr
{\rm Along}\;x_1\; {\rm direction}\;&& V_{\rm eff}\sim\left\{\ba{ll} -\ell^{\frac{2\theta-d}{d}}& \theta\neq \frac{d}{2},\cr &\cr -\ln\frac{\ell}{\epsilon}&\theta= \frac{d}{2}.\ea\right.
\eea
It is worth to note that since $\theta\leq d$, it is possible (for example for $\theta=d$) to get linear or 
fractional power law  effective potential. In fact due to the anisotropic in the spatial directions, it 
is possible to get a power law effective potential in  one direction while
a confining potential in other directions.

\subsection{Drag force}

Let us consider an open string moving in the background \eqref{metric1}  
 which 
could represent an external  point like  moving source in an effective theory with hyperscaling 
violation. This can be used to study the drag force that the external object 
might feel\cite{Gubser:2006bz}\footnote{ After we submitted our paper we were aware that the
drag force in the background \eqref{metric1} has also been, recently, sudied in \cite{Kiritsis:2012ta}.} .
Drag force  for non-relativistic theories with Schr\"odinger or Lifshitz symmetries have 
been studied in \cite{Akhavan:2008ep} and \cite{Fadafan:2009an}, respectively, where it was 
shown that even at zero  temperature the drag force is non-zero. It is indeed the aim of this 
subsection to study the drag force for theories with hyperscaling violation.

Consider  the following  ansatz for the moving string
\be
t=\tau,\;\;\;\;\;\;r=\sigma,\;\;\;\;\;\;x_1=vt+x(r),
\ee
and all other coordinates are set to zero. 
Then the Nambu-Goto action for this string reads
\be
I=-\frac{1}{2\pi\alpha'}\int dt dr\; r^{2\frac{\theta}{d}-2}\sqrt{r^{2(1-z)}(1+{x'}^2)-v^2},
\ee
 Since the metric components
are $t$ independent, the above action may be treated as a one dimensional mechanical
system whose momentum is the constant of motion
\be
 r^{2\frac{\theta}{d}-2}\;\frac{r^{2(1-z)}{x'}}{\sqrt{r^{2(1-z)}(1+{x'}^2)-v^2}}=-2\pi\alpha' \Pi_x=
{\rm constant},
\ee
which can be solved to find $x'$ as follows
\be\label{cons}
{x'}^2=4\pi^2{\alpha'}^2\Pi_x^2\;\frac{1-r^{2(z-1)}v^2}{r^{4\frac{\theta}{d}-2(z+1)}-
4\pi{\alpha'}^2\Pi_x^2}.
\ee
In terms of the constant $\Pi_x$,  energy, $E$, and  momentum, $P$, that the  open string gains from  
through its end point are given by 
\be 
\frac{dE}{dt}=\Pi_x v.\;\;\;\;\;\;\;\;\;\;\;\frac{dP}{dt}=\Pi_x.
\ee
The constant $\Pi_x$ can be obtained by requiring
that the numerator and denominator of \eqref{cons} vanish at the same point,
imposed by the fact that $x'$ should be real. Setting the numerator  to zero one finds $v=r_0^{1-z}$. 
From this solution we  observe
that as we are varying $r_0$ from infinity to zero (flowing from IR to UV), the velocity takes its value from
 zero to infinity.
This is, in fact, due to the non-relativistic property of the dual field
theory. Moreover by  plugging the solution $r_0$ into the denominator one arrives at
\be
\Pi_x=-\frac{v}{2\pi {\alpha'}}\; v^{\frac{2(d-\theta)}{d(z-1)}}.
\ee
Now consider a single non-relativistic particle with momentum $P$ and mass $M$,
then we have $P = Mv$. It is useful to formally rewrite the above expression for the
drag force in terms of $P$. Then we can perform the integral which for $d\neq \theta$ yielding
to
\be
P=\left(P_0^{-\frac{2(d-\theta)}{d(z-1)}}+\frac{d-\theta}{d(z-1)\pi\alpha'}\;
\frac{t}{M^{\frac{d(z-3)+2\theta)}{d(z-1)}}}\right)^{-\frac{d(z-1)}{2(d-\theta)}},
\ee
while for $d=\theta$ one finds
\be
P=P_0e^{-\frac{t}{2\pi\alpha' M}}.
\ee
Here $P_0$ is the initial momentum.

We note that even though  the theory we considered here was at zero
temperature the drag force is non-zero. In other words, as the open string moves on the metric \eqref{metric1} it experiences a non-zero
friction. Therefore if we let the open string (external point like object) moves with an initial velocity (initial momentum $p_0$)  after some time it will stop unless one compensates the losing energies by an 
external force. 

Actually having a non-zero drag force even at zero temperature appears when there is  an 
anisotropic scaling between time and the direction the string moves. This is of course a typical 
feature of non-relativistic
field theories whose gravitational duals have metrics with an anisotropic scaling.

In fact it is straightforward to compute the induced metric on the string  worldsheet and  argue 
that there is an event. For example for an open string moving in the geometry \eqref{metric1} the induced metric on its worldsheet is
\be
ds^2=r^{\frac{2\theta}{2}-2}\left[-(r^{2(1-z)}-v^2)dt^2+(1+{x'}^2)dr^2+2vx'dr dt\right],
\ee
which can be diagonalized by the following change of coordinates\cite{Gursoy:2010aa}
\be
t=\tau+\eta(r),\;\;\;\;\;\;{\rm with}\;\;\eta'(r)=\frac{v x'}{r^{2(1-z)}-v^2},
\ee
by which the induced metric reads
\be
ds^2=r^{\frac{2\theta}{2}-2}\left[-(r^{2(1-z)}-v^2)d\tau^2+\frac{r^{\frac{4\theta}{d}-2(z+1)}}{r^{\frac{4\theta}{d}-2(z+1)}-4\pi^2\alpha'^2\Pi_x^2} dr^2\right].
\ee
As we see the induced metric develops an event horizon whose location is given by $r_0^{2(1-z)}-v^2=0$, and the Hawking temperature is
\be
T=\frac{\sqrt{(z-1)(z+1-\frac{2\theta}{d})}}{2\pi r_0^z}.
\ee
On observes that even the bulk is at zero temperature, the worldsheet is thermal which should be 
responsible of having energy loss in this case. Note that for  $z=1$ the temperature is zero and 
indeed for this the energy loss is zero too.

Let us now study an open string  moving  in the metric \eqref{metric2}. In this case we 
could consider a situation where either the string moves in $x_1$ or $y$ directions. In the first case,
actually, due to the rotational symmetry in $x_1$ and $t$ directions when the string moves in $x_1$ direction, one gets nothing new, except 
that of the pure AdS case\cite{Gubser:2006bz}. Namely there is no drag force.

On the other hand when the open string moves in $y$ direction, the induced metric on its
worldvolume is
\be
ds^2=r^{\frac{2\theta}{2}-2}\left[-(1-v^2r^{2(1-z)})dt^2+(1+{x'}^2r^{2(1-z)})dr^2+2vx'r^{2(1-z)}dr dt\right],
\ee
which, by making use of the equation of motion of $\phi$, can be recast to the following form
\be
ds^2=r^{\frac{2\theta}{2}-2}\left[-(1-v^2r^{2(1-z)})d\tau^2+\frac{r^{\frac{4\theta}{d}-2(z+1)}}{r^{\frac{4\theta}{d}-2(z+1)}-4\pi^2\alpha'^2\Pi_x^2} dr^2\right].
\ee
It seems that the resultant world-volume metric has an event horizon and the situation could be the 
same as before. We note, however, that if we went through the standard procedure, as we did in the 
previous case, the Hawking temperature turns out to be imaginary which could mean that the
ansatz we considered for the  open string is not a consistent solution. Therefore it seems that even in this 
case where the string moves along a direction with an anisotrpoic scaling, the drag force is zero.

\subsection{Rotating open string}

As a final example of open strings probing the geometries \eqref{metric1} and \eqref{metric2},  in this subsection we will consider a rotating open string in the bulk that describes, holographically, an 
external point like object in the strongly coupled theories with hyperscaling violation undergoes
circular motion. Such a study  has been first done for a rotating quark in strongly coupled 
${\cal N}=4$ SYM theory in\cite{{Fadafan:2008bq},{Athanasiou:2010pv}} where the authors ``have  computed the energy 
density and angular distribution of the power radiated by a rotating quark in this theory.''\footnote{
Rotating massive quark in an anisotropic strongly coupled plasma has also 
been studied\cite{Fadafan:2012qu}. 
For rotating strings on  non-conformal holography see \cite{AliAkbari:2011ue}.}
To start let us  consider the following change of coordinates in the metrics \eqref{metric1} and \eqref{metric2}
\be
(x_1,x_2)\rightarrow (\rho,\phi);\;\;\;\;\;\;{\rm such\; that}\;\;\;dx_1^2+dx_2^2=d\rho^2+\rho^2d\phi^2,
\ee
by which the metrics \eqref{metric1} and \eqref{metric2} may be recast to the following forms
\bea\label{met1}
&&ds^2_{d+2}=r^{\frac{2\theta}{d}}\left(-\frac{dt^2}{r^{2z}}+\frac{dr^2}{r^2}+\frac{d\rho^2}{r^2}+\frac{\rho^2 d\phi^2}{r^2}+\frac{\sum_{i=3}^ddx_i^2}{r^2} \right),\\
&&ds_{d+2}^2=r^{\frac{2\theta}{d}} \left(\frac{dy^2}{r^{2z}}+\frac{dr^2}{r^2}+
\frac{d\rho^2}{r^2}+\frac{\rho^2d\phi^2}{r^2}+\frac{\sum_{i=3}^{d-1}dx_i^2
}{r^2}-\frac{dt^2}{r^2} \right).\label{met2}
\eea
Let us now consider the following ansatz for a rotating open string in these backgrounds 
\be
t=\tau,\;\;\;\;\;r=\sigma,\;\;\;\;\;\;\rho=\rho(r),\;\;\;\;\;\;\;\phi=\omega t+\phi(r),
\ee
and all other coordinates are set to zero. From dual field theory point of view this corresponds to an extremal point like object undergoes a circular motion  around the center $\rho=0$ with frequency $\omega$. 

The Nambu-Goto action for this open string in the background \eqref{met1} is
\be\label{act1}
I=-\frac{1}{2\pi\alpha'}\int dt dr \; r^{2\frac{\theta}{d}-1-z}\sqrt{(1-\rho^2\omega^2r^{2(z-1)})(1+{\rho'}^2)+\rho^2{\phi'}^2},
\ee
whereas for the metric \eqref{met2} it becomes
\be\label{act2}
I=-\frac{1}{2\pi\alpha'}\int dt dr \; r^{2\frac{\theta}{d}-2}\sqrt{(1-\rho^2\omega^2)(1+{\rho'}^2)+\rho^2{\phi'}^2}.
\ee
Note that the action \eqref{act2} up to a factor of $r^{2\frac{\theta}{d}}$ is the same as that 
for pure AdS case studied in \cite{Athanasiou:2010pv} .

In both cases the actions are  independent of $\phi(r)$, and therefore the corresponding angular momentum,
$\Pi_\phi=-\frac{\partial{\cal L}}{\partial\phi'}$,  is an integral of motion. In other words for the action \eqref{act1}, setting $p=2\frac{\theta}{d}-1-z$ and $q=2(z-1)$ one has
\be
\Pi_\phi =\frac{-r^p\rho^2 \phi'}{\sqrt{(1-\rho^2\omega^2r^q)(1+{\rho'}^2)+\rho^2{\phi'}^2}} ={\rm constant }, 
\ee
while for action \eqref{act2}, setting $\tilde{p}=2\frac{\theta}{d}-2$ one gets
\be
\Pi_\phi =\frac{-r^{\tilde{p}}\rho^2 \phi'}{\sqrt{(1-\rho^2\omega^2)(1+{\rho'}^2)+\rho^2{\phi'}^2}} ={\rm constant }.
\ee
Using these expressions one can find the  equation of motion for $\phi(r)$ as follows
\bea\label{phi1}
{\rm for\; action\; \eqref{act1}} &&\phi'^2=\frac{\Pi_\phi^2 (1-\omega^2 \rho^2r^q) (1+\rho'^2)}{\rho^2(r^{2p}\rho^2-\Pi_\phi^2)},\\
{\rm for\; action\; \eqref{act2}} &&\phi'^2=\frac{\Pi_\phi^2 (1-\omega^2 \rho^2) (1+\rho'^2)}{\rho^2(r^{2\tilde{p}}\rho^2-\Pi_\phi^2)}.\label{phi2}
\eea
Finally, utilizing the above expressions,  the equation of  motion for $\rho(r)$,  is then given by
\bea
{\rm for\; action\; \eqref{act1}} &&\rho''+\left(\frac{\rho (r-p\rho\rho')}{r(r^{-2p}\Pi_\phi^2-\rho^2)}+
\frac{(2-qr^{q-1}\omega^2\rho^3\rho')}{2\rho(1-\rho^2\omega^2r^q)}
\right)(1+\rho'^2)=0,\cr &&\cr
{\rm for\; action\; \eqref{act2}} &&\rho''+\left(\frac{\rho (r-\tilde{p}\rho\rho')}{r(r^{-2\tilde{p}}\Pi_\phi^2-\rho^2)}+\frac{1}{\rho(1-\rho^2\omega^2)}
\right)(1+\rho'^2)=0.
\eea
We have now all equations to study the rotating strings. To proceed, in what follows we will only  consider
the first case which is based on the action \eqref{act1}.  Actually the results for the second case  can be obtained from the first one by replacing $p\rightarrow \tilde{p}$ and $q\rightarrow 0$.

In general to find a solution for the rotating string we have to solve the equations of motion of $\rho$ and $\phi$
with a given initial data\cite{Athanasiou:2010pv}. Actually to solve the equation of motion for $\rho$ we find that it is 
singular when $\Pi_\phi^2-\rho^2r^{2p}=0$ or  $1-\rho^2\omega^2r^q=0$. On the 
other hand in order to impose the reality condition on $\phi'$ one has to assume that 
these two expressions should vanish at the same point. Let us denote by $r_0$ the point where 
both of the expressions vanish. Then one gets
\be\label{rho0}
\rho_0=\frac{(\Pi_\phi\omega)^{\frac{-q}{2p-q}}}{\omega},\;\;\;\;\;\;\;\;\;\;r_0=(\Pi_\phi\omega)^{\frac{2}{2p-q}},
\ee
where $\rho_0=\rho(r_0)$. On the other hand $\rho'(r_0)$ may also be fixed by $\omega$ and $\Pi_\phi$
as follows. Following \cite{Athanasiou:2010pv} let us expand $\rho$ around $r_0$,
\be
\rho=\rho_0+\rho_1(r-r_0)+\frac{1}{2}\rho_2(r-r_0)^2+\cdots.
\ee
Plugging this expansion in the equation of motion for $\rho$ and requiring all terms in the expansion vanish identically we can read the coefficients $\rho_i$.  At the first order, for $2p+q\neq 0$, one finds
\be
\rho_1^2 - \frac{2\left(2\omega^2-pq(\Pi_\phi \omega)^{\frac{2q+4}{q-2p}}\right)}{(q+2p)\omega (\Pi_\phi \omega)^{\frac{q+2}{q-2p}} } \rho_1-1=0,
\ee  
which can be solved to find $\rho_1$ in terms of $\omega$ and $\Pi_\phi$. 

 Therefore with $\omega$ and $\Pi_\phi$ as the initial data one should be able to find a unique solution
for the rotating open string. Indeed, by making use of a numerical method we have solved the equation of motion for $\rho$ with specific
 values of $p$ and $q$. In particular in the figure (1) we have plotted
$\rho$ as a function of $r$ for the cases of $q=2$ and $p=-3/2,-2,-3$ which correspond to 
the cases where $z=2$ and $\theta= \frac{3d}{4},\frac{d}{2},0$. 
In order to compare the results with that in the pure AdS case\cite{Athanasiou:2010pv}, we have also plotted $\rho$ as a function of 
$r$ for the case of $z=1,\theta=0$.
\begin{figure}
\begin{center}
\includegraphics[scale=0.16]{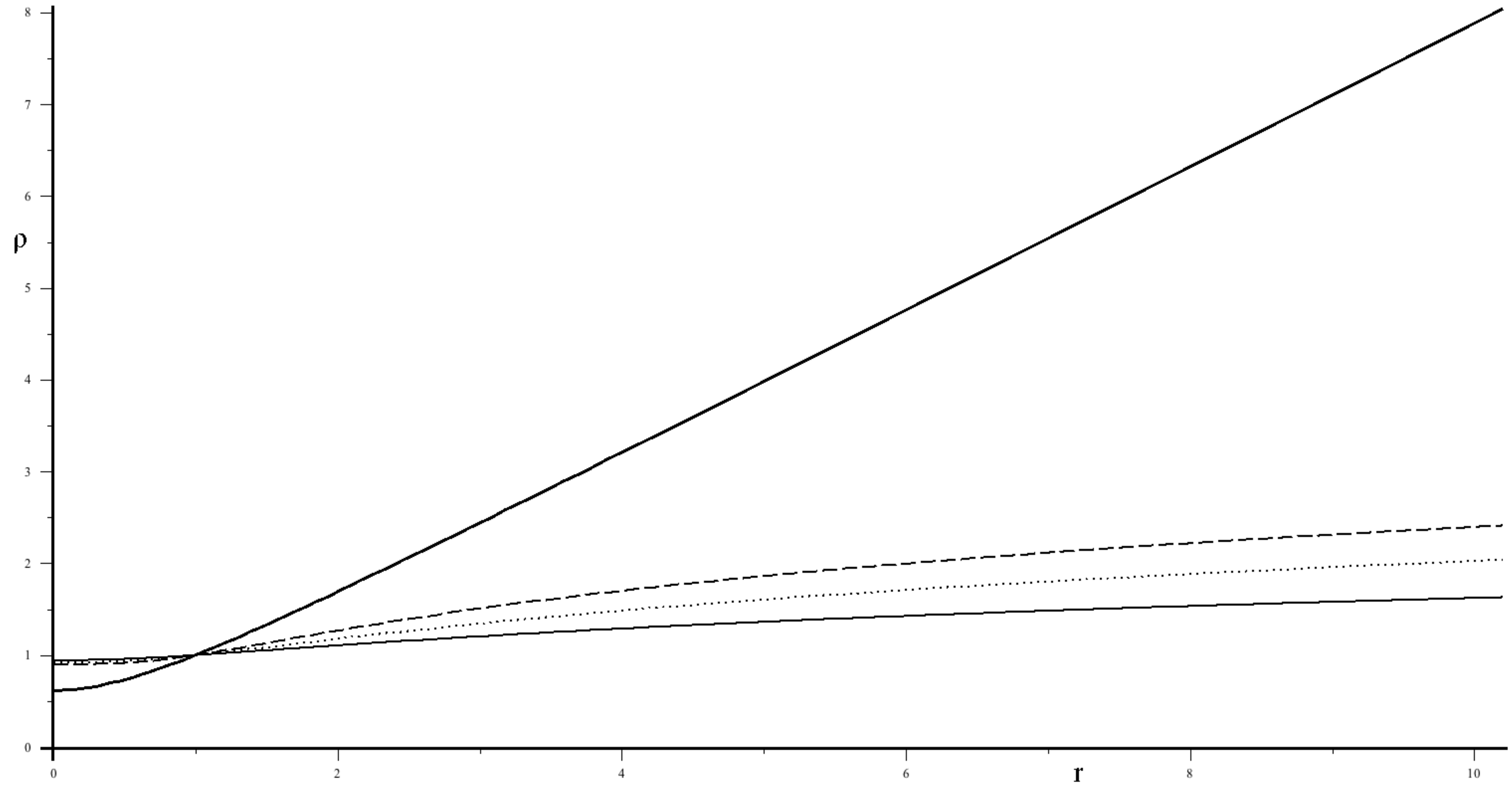}
\caption{$\rho$ as a function of $r$ for the cases of  $z=2$ and $\theta=\frac{3d}{4},\frac{d}{2},0$ which 
are shown by  dashed, dotted and thin curves, respectively. In order to compare our 
results with that of  AdS case ($z=1, \theta=0$), we have also plotted this case which is shown by a
thick curve.   Note that to solve the equation we 
have set the initial date as $\omega=\Pi_\phi=1$.}
\end{center}\label{fig1}
\end{figure}

The function of  $\rho(r)$ shows how the string spreads as a function depth\cite{Athanasiou:2010pv}.
Actually as we can see from the figure (1) for the rotating string in the AdS case, the string spreads linearly, $\rho\sim r$, as we approach the IR region (see also the exact solution given in \cite{Athanasiou:2010pv}). On the other hand although in the present  case the string still 
spreads  as a function of depth, the rate of spreading is slower than that in the AdS case. In fact in our
case, as one can see from our numerical solution, the asymptotic behavior is $\rho\sim r^\alpha$
for $r\gg 1$ with $\alpha< 1$. Of course for $r<1$ the string on the hyperscaling violation
metrics spreads faster than that on the pure AdS (see figure 1). Pictorially, these 
behaviors can be visualized by projecting the function $\rho(\phi)$ to a $r={\rm constant}$ (boundary) surface as we have depicted in figure (2).
\begin{figure}
\begin{center}
\includegraphics[scale=.05]{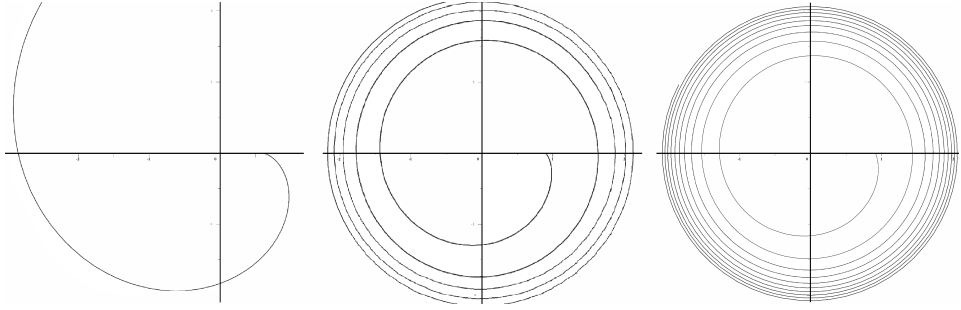}
\caption{Here we have plotted $\rho$ as a function of $\phi$ for the cases of  $z=1, \theta=0$ (left), $z=2, \theta=0$ (middle) and $z=2, \theta=
\frac{d}{2}$ (right). The function is projected in to $x_1-x_2$ plane, where $\rho$ and $\phi$ are polar coordinates of this place. Here we have set
$\omega=\Pi_\phi=1$.}
\end{center}
\end{figure}

On the other hand looking at the behavior of the function $\phi$ which describes how the string
 winds in azimuthal angle as a function of depth\cite{Athanasiou:2010pv}, one finds that 
in the present case where we have the hyperscaling violation, it grows faster that that in AdS as we move
toward the core of  the space-time.

For $2p+q=0$, expanding $\rho$ around $\rho_0$ we find that all $\rho_i=0$. Indeed it is easy to show that in this case $\rho={\rm constant}$ is an exact 
solution of the equation of motion where the constant is found to be 
$\rho=\sqrt{\Pi_\phi/\omega}$ which is consistent with \eqref{rho0}. Moreover, 
in this case the equation of motion of $\phi$ can also be solved exactly.  In fact for $2p+q=0$, which corresponds to $\theta=d$ for all values of $z$, one finds\footnote{Note that 
this solution is not valid for $z=1$. Actually for $\theta=d$ and $z=1$ the geometry reduces to 
the flat space-time where we do not have a well defined holography.}
\be
\phi=\omega\left(t+\frac{r^z}{z}\right),\;\;\;\;\;\;\;\;\;\;\;\rho=\rho_0=\sqrt{\frac{\Pi_\phi}{\omega}},\;\;\;\;\;\;\;\;\;\;r_0=(\Pi_\phi\omega)^{\frac{1}{2(1-z)}}.
\ee

To wrap up this section we note that  looking at the induced metric of the string worldsheet one
 observes that it has an even horizon. More precisely, with a proper change of coordinates,  the 
induced metric of the string worldsheet is found
\be
ds^2=r^{\frac{2\theta}{d}-2-q}\left[-(1-\omega^2\rho^2r^q)d\tau^2+\frac{r^{2p+q}\rho^2(1+\rho'^2)}
{r^{2p}\rho^2-\Pi_\phi^2}dr^2\right],
\ee
which has an even horizon at $r=r_0$. The corresponding  Hawking temperature is 
\be
T=\frac{1}{4\pi \rho_0 r_0^z}\sqrt{-\frac{2(q\rho_0+2\rho_1r0)(p\rho_0+\rho_1r_0)}{ 1+\rho_1^2}}.
\ee
It is crucial to note that to have a solution for $\rho(r)$ in the vicinity of $r_0$, the positivity condition of the right hand side of the equation (\ref{phi1}) reduces to the following inequality
\be
(q\rho_0+2\rho_1r0)(p\rho_0+\rho_1r_0)<0,
\ee
which is necessary to have a well defined real Hawking temperature.

Having found the shape of the rotating  string, it is then possible to compute the radiation power 
of the string. From the Nambu-Goto action one can obtain the  energy density and energy flux
of string from which  the energy loss rate of the  string is found\cite{Athanasiou:2010pv}
\be
\frac{dE}{dt}=\frac{\Pi_\phi \omega}{2\pi\alpha'}.
\ee
Of course the goal is to compute the energy loss rate in terms of the parameters of the dual field 
theory which is frequency and the radius of circle the external object moves around, {\it i.e.} $\rho_*=\rho(r\rightarrow 0)$. With the numerical solution we have obtained, it is easy to find $L=\Pi_\phi\omega$
as a function of $\rho_*$ which is indeed the energy loss rate up to factor of $1/2\pi\alpha'$. 
In the figure (3) we have drown the function of $L$ in terms of $\rho_*$ for the case of $z=2,\theta=0$. 
\begin{figure}
\begin{center}
\includegraphics[scale=.2]{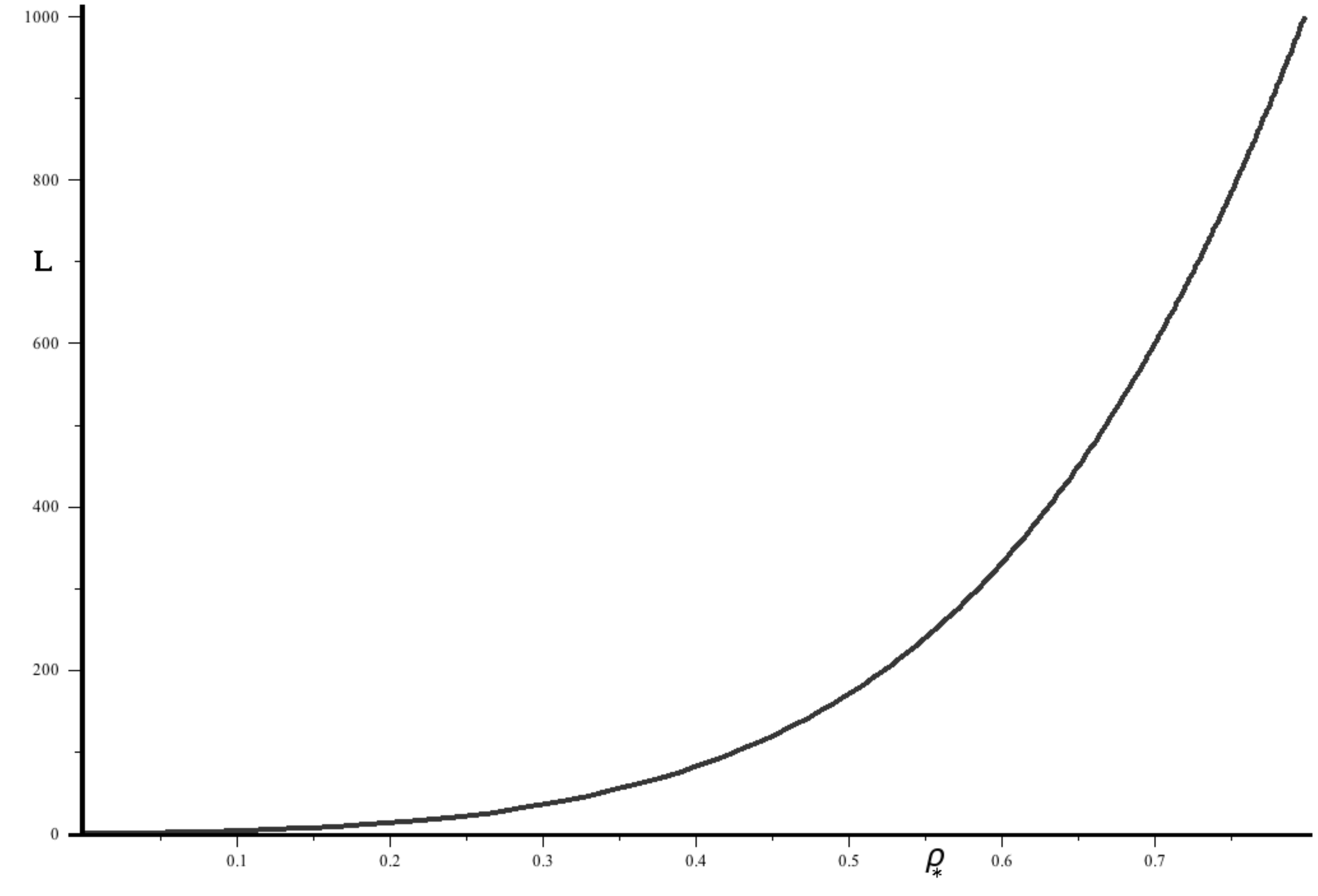}
\caption{Here we have plotted $\rho_{*}$ as a function of $L=\Pi_\phi\omega$ for the case of  $z=2, \theta=0$. Essentially this is the energy loss rate in terms of the radius of circle the external object moves.}
\end{center}
\end{figure}
Form our numerical solution, for large $\rho_*$, one finds $L\sim \rho_*^4$.

As a final remark we note that  $r_0$ fixes the local velocity of string. Since 
$r_0$ can change from zero to infinity, it means that the local velocity could be infinity, indicating 
the non-relativistic nature  of the dual theory.


\section{Closed string probe}

Since it is possible to find a geometry with non-trivial hyperscaling exponent from string theory
(see for example \cite{{Dong:2012se},{Narayan},{AMK}}) it is natural to study a closed string propagating in this background
which could be dual to an operator in the dual field theory with n hyperscaling violation. In fact 
in this section, for completeness of our study, we will  probe the geometry \eqref{metric1} by a closed string. Via AdS/CFT correspondence a closed string moving in the
bulk should correspond to a field (operator) configuration with specific energy and probably some other quantum numbers. Therefore it could be possible to study certain 
properties of the dual theory by making use of a semi-classical closed string on the bulk. 
In what follows we will  consider a folded rotating closed\cite{Gubser:2002} and circular pulsating\cite{Minahan:2002rc}
 strings in the background \eqref{metric1}.\footnote{Circular pulsating closed string on the 
confining geometry has also been studied in \cite{mosaffa}.}

\subsection{Folded closed string}

In this subsection we study a solution representing a rotating 
closed string configuration which is stretched along the 
radial coordinate. In order to study this system one needs 
to write an action for this closed string.
Let us parameterize the string worldsheet by $\sigma $ and $\tau$. We can fix the reparameterization 
invariance by a parameterization such that the time coordinate of space-time, 
$t$ to be equal to worldsheet time, i.e. $t=\tau$. 
In this gauge a closed string configuration
representing a rotating string with angular velocity 
$\omega$ on geometry (\ref{metric1}) stretched along the radial 
coordinate is given by  
\bea
&& t=\tau,\qquad  \phi_{d-1}=\phi=\omega \tau,\qquad r(\sigma)=r(\sigma+2\pi),\cr\cr 
&&  \phi_i=\frac{\pi}{2} \;\;({\mathrm for} \; i=1\cdots d-2)  ,\quad  \rho=\rho_0 ={\rm constant.} 
\label{CL}
\eea
where $\sigma$ and $\tau$ are coordinates of the string world sheet. Moreover we parametrize the $d$-dimensional Euclidean space in the metric (\ref{metric1}) as follows
\be
\sum_{i=1}^d dx_i^2=d\rho^2+\rho^2 (d\phi_1^2+\sin^2\phi_1d\phi_2^2+\cdots+\sin^2\phi_1...\sin^2\phi_{d-2}d\phi_{d-1}^2).
\ee
For this  ansatz  the Nambu-Goto action,
reads
\be
I=-4\frac{T}{2\pi\alpha' }\int_{0}^{r_0} r^{\frac{2\theta}{d}-2}\sqrt{r^{2-2z}\dot{t}^2-\rho_0^2\dot{\phi}^2}\; dr
\ee
where $dot$ represents derivative with respect to $\tau$ and $r_0=(\rho_0\omega)^{1/(1-z)}$. The factor of 4 comes from the fact that we are dealing with a
folded closed string. Working with one fold string, the string can be divided to four
segments. Using the periodicity condition we just need to perform the integral for
one quarter of string multiplied by factor 4.  When the periodicity condition is satisfied, using the above Nambu-Goto action the conserved quantities are given by
\bea
 E=\frac{2T}{\pi\alpha'}\int_{0}^{r_{0}}   dr \frac{ r^{\frac{2\theta}{d}-2z}}{     \sqrt{r^{2-2z}-\rho_0^2\omega^2}}, \;\;\;\;\;\;\;\;S=\frac{2T\rho_0^2\omega}{\pi\alpha' }
\int_{0}^{r_{0}}   dr \frac{ r^{\frac{2\theta}{d}-2}}{     \sqrt{r^{2-2z}-\rho_0^2\omega^2}},
\eea
From these expressions we could find a relation between $E$ and $S$. Actually the above integrals can be performed exactly, though generically they 
diverge and need to be regularized.  Indeed
setting $x=r/r_0$, one has
\bea\label{En-Sp}
&&{ \cal E}=\frac{\pi\alpha'}{2T}E=r_0^{\frac{2\theta}{d} -z}   \int_0^{1}\frac{ x^{\frac{2\theta}{d}-z-1}}{\sqrt{1-x^{2z-2}}}\; dx ={\cal E}_0  r_0^{\frac{2\theta}{d} -z}\\ 
&& {\cal S}= \frac{\pi\alpha'}{2T\rho_0}S=r_0^{\frac{2\theta}{d} -1} \int_0^{1}\frac{ x^{\frac{2\theta}{d}+z-3}}{\sqrt{1-x^{2z-2}}}\; dx= {\cal S}_0 r_0^{\frac{2\theta}{d} -1} \eea  
where constants ${\cal E}_0$ and ${\cal S}_0$ are given by
\be
{\cal E}_0=\frac{\sqrt{\pi}}{2(z-1)}\frac{\Gamma\left(\frac{2\theta-zd}{2d(z-1)} \right)}{\Gamma\left(\frac{2\theta-1}{2d(z-1)} \right)},\quad   {\cal S}_0=\frac{\sqrt{\pi}}{2(z-1)}\frac{\Gamma\left(\frac{2\theta+zd-2d}{2d(z-1)} \right)}{\Gamma\left(\frac{2\theta+2zd-3d}{2d(z-1)} \right)}.
\ee
Therefore with a proper regularization one finds
\be\label{ES}
{\cal E}={\cal E}_0{\cal S}_0^{\frac{zd-2\theta}{2\theta-d}}\; {\cal S}^{\frac{2\theta-zd}{2\theta-d}}.
\ee

Looking at the explicit form of the energy and spin,  depending on the nature of the divergences  we recognize two distinct cases. Actually although in a typical case we get 
power law divergence, there are cases where we get logarithmic divergences that correspond to the cases where 
either $2\theta= zd$, or $\theta=0,\;z=2$. Note that the latter case is indeed Lifshitz geometry with scaling exponent $z=2$. 

For  $2\theta= zd$ the energy diverges  due to UV divergence of the integral. Therefore we need to regularize it by performing the integral with a 
UV cut off $\epsilon$. Doing so, one arrives at  
\be
{\cal E}=\int_\epsilon^{r_0}\frac{dr}{r\sqrt{1-(\frac{r}{r_0})^{2(z-1)}}}=\ln \frac{r_0}{\epsilon}+\frac{1}{z-1}\ln2.
\ee
On the other hand in this case one has ${\cal S}=\frac{r^{z-1}}{z-1}$. Therefore one finds
\be
{\cal E}=\frac{1}{z-1}\ln \frac{2(z-1)\cal S}{\epsilon^{z-1}}.
\ee

In the case of  $\theta=0,\;z=2$, although the energy and spin both diverge
logarithmically, the relation between energy and spin  is still given by the equation \eqref{ES} for 
$\theta=0$ and $z=2$.

It is also obvious  from the equation \eqref{ES} that the expression becomes ill-defined for 
$\theta=\frac{d}{2}$. In fact in this case  starting from the  original expression for the energy and spin
in the equation \eqref{En-Sp} one finds that the spin is a constant for any turning point, {\it i.e.} 
${\cal S}=\frac{\pi}{2(z-1)}$, while the energy is zero.

It is interesting to note that although the closed string, depending on $\omega$  could be short or long, the relation between energy and spin remains unchanged in all
cases discussed above. It is 
in contrast to the AdS case where for short string ($\omega\rightarrow  \infty$) we find  the 
flat space Regge trajectory, while for long string ($\omega\rightarrow 0$)  the energy in terms 
of spin has logarithmic  behavior\cite{Gubser:2002}.

\subsection{Circular pulsating string}

In this subsection we will consider a pulsating string which wrapped $m$ times around $\phi_{d-1}$
direction. In other words we study a circular string that expand and contract in 
the radial direction. The corresponding ansatz for the circular pulsating string is
\be
t=\tau,\;\;\;\;\;\;\;r=r(t),\;\;\;\;\;\;\;\phi_{d-1}=m\sigma,\;\;\;\;\;\;\;\;\phi_i=\frac{\pi}{2},\;\;\;\;\;\;\;\;\;\rho=\rho_0={\rm constant}.
\ee
 For this configuration the string action reads
\be
I=-\frac{m\rho_0}{\alpha'}\int dt\; r^{2\frac{\theta}{d}-z-1}\sqrt{1-r^{2(z-1)}\dot{r}^2}.
\ee
Setting $\xi=r^z/z$ the above action can be recast to the following form
\be
I=-\frac{m\rho_0}{\alpha'}\int dt\; f(\xi) \sqrt{1-\dot{\xi}^2},\;\;\;\;\;\;\;{\rm with}\;f(\xi)=(z\xi)^{\frac{2\theta-zd-d}{zd}}.
\ee
This action can be treated as a one-dimensional quantum mechanical system whose Hamiltonian is 
\be
H=\sqrt{\Pi^2+(\frac{m\rho_0}{\alpha'})^2f^2(\xi)}
\ee
 where $\Pi$ is the canonical momentum  given by
\be
\Pi=\frac{m\rho_0}{\alpha'}f(\xi)\;\frac{\dot{\xi}}{\sqrt{1-\dot{\xi}^2}}.
\ee
Following \cite{Minahan:2002rc} $H^2$ can be considered as a one dimensional quantum mechanical system with the potential $(\frac{m\rho_0}{\alpha'})^2f^2(\xi)$. Therefore
using  the Bohr-Sommerfeld quantization we have
\be
(n+\frac{1}{2})\pi=E\int_{\xi_0}^\infty d\xi\;\sqrt{1-\left(\frac{ f(\xi)}{f(\xi_0)}\right)^2},
\ee
where $\xi_0$ defined by the root of $f(\xi_0)=\frac{\alpha' E}{m\rho_0}$. Performing the integral, with a proper regularization, one finds
\be 
(n+\frac{1}{2})\pi=-\frac{\sqrt{\pi}}{2z}\;\frac{\Gamma\left(\frac{d-2\theta}{d(z+1)-2\theta}\right)}
{\Gamma\left(\frac{1}{2}+\frac{d-2\theta}{d(z+1)-2\theta}\right)} \;
\left(\frac{\alpha' E}{m\rho_0}\right)^{\frac{zd}{2\theta-zd-d}}E.
\ee

It is worth to note that, unlike the AdS case,  in the present  case due to the non-trivial IR geometry, the string while pulsating cannot approach $r=0$. Indeed there is a low bound for the string in the
radial direction fixed by 
\be
r\ge \left(\frac{m\rho_0}{E\alpha'}\right)^{\frac{d}{d+dz-2\theta}}.
\ee

\section{Discussions   }

In this paper we have studied certain features of strongly coupled theories with hyperscaling violation by making use of AdS/CFT correspondence. More precisely we have considered a gravitational model 
on a background which is conformally a metric with the Lifshitz isometry. The metric has two 
parameters corresponding to dynamical and hyperscaling violation exponents. This geometry may be found
from  an Einstein-Maxwell-Dilaton theory.

By making use of a double Wick rotation it is possible to have a situation where the anisotropic scaling
could be  along either  time or one of spatial directions.  In both cases, using AdS/CFT correspondence,
 we have computed
several quantities including holographic entanglement entropy,
Wilson loop, drag force as well as energy loss of a rotating string. 

 Although we have studied the different quantities for a generic dynamical and hyperscaling violation exponents in arbitrary dimensions, we have found that for  particular values  of the parameters
the theory exhibits interesting behavior. 

To compute the entanglement entropy for a strip in a theory with an anisotropic scaling
along a spatial direction we considered two distinctive cases depending  on
whether the width of the strip is along the anisotropic direction. When 
the width of the strip is along the anisotropic direction we have found that when 
$\theta=d-1$ for any $z$, the entanglement entropy exhibits a logarithmic violation of area
law while for $\theta=d$, the system has an extensive entropy. Indeed the situation is very
similar to the case where the anisotropic scaling is along the time direction\cite{Dong:2012se}.
On the other hand for the case where the width of strip is along a spatial direction with a normal 
scaling, the entanglement entropy shows logarithmic behavior for $\theta=d$ and $z=2$, while in 
this case the subsystem never has an extensive  entropy. Having found the logarithmic behavior 
might  indicate that the model has a Fermi surface.

For an open string probing the geometry one can read several interesting quantities, such as Wilson loop 
which in turn can be used to read the effective potential between two external point like 
objects in the system. We have shown that typically the effective potential is power low, though
for special cases it could be logarithmic. 
  
For a moving or rotating open string we have observed  that even though the geometry are at
zero temperature, the induced metric on the worldvolume of the open string has an event 
horizon leading to a non-zero Hawking termperature for worldvoulme theory. This might be 
the reason we are getting non-zero drag force for the string. Again in this case we find that
$\theta=d$ is an special case where the string shows an anomalous behavior.

We have also probed the background with different closed strings. The closed strings we have 
considered include rotating folded and circular pulsating strings. From the gravity point of view 
it is possible to find the energy of the strings in terms of the quantum numbers the strings may have.
This may be used to find  the dispersion relation of a possible field configuration in the dual strongly 
coupled hyperscaling violation field theory. The anomalous behavior can also been seen for 
closed string too. In particular for $zd=2\theta$ ( which includes $\theta=d,z=2$) the 
dispersion relation is logarithmic.

In this paper we have only considered a zero temperature background, though it could be generalized
to the finite temperature too. In fact the finite temperature metric has the following form\cite{Dong:2012se}
\be
ds_{d+2}^2=\left( \frac{r}{r_F}\right)^{\frac{2\theta}{d}}\left(-r^{-2(z-1)}f(r)dt^2+\frac{dr^2}{f(r)} + dx_i^2\right),
\ee 
where 
\be
f(r)=1-\frac{r^{d+z-\theta}}{r_h^{d+z-\theta}} 
\ee   
It would be interesting to study the quantities we have considered in this paper at  finite temperature.

\vspace*{1cm}

{\bf Acknowledgments}

We would like to thank Eoin \'O Colg\'ain, Mohammad  Mohammadi Mozaffar, Ali Mollabashi, Amir E. Mosaffa and Ali Vahedi for discussions on 
different aspects of holographic entanglement entropy and hyperscaling violation theories.
 The research of M.A. is supported by Iran National Science Foundation (INSF). The work of H.Y.  is supported by the National Research Foundation of Korea Grant funded by the Korean Government (NRF-2011- 0023230)

\end{document}